\documentclass[prc,unsortedaddress,superscriptaddress,twocolumn,amssymb,showpacs,floatfix,nofootinbib]{revtex4}
\usepackage{graphics,epsfig}
\usepackage{amsmath}
\usepackage{dcolumn}   % needed for some tables
\begin{document}
\date{\today}
\title{Resonance-state properties from a phase shift analysis with the $S$-matrix pole method and the effective-range method}
\author{B.F. Irgaziev}\email{irgaziev@yahoo.com}
\affiliation{Institute of Applied Physics, National University of
Uzbekistan} \affiliation{GIK Institute of Engineering Sciences and
Technology, Topi, Pakistan}
\author{Yu.V. Orlov}\email{orlov@srd.sinp.msu.ru}
\affiliation{Skobeltsyn Nuclear Physics Institute, Lomonosov Moscow State
University, Russia}

\begin{abstract}
Asymptotic normalization coefficients (ANCs) are fundamental
nuclear constants playing an important role in nuclear physics and
astrophysics. We derive a new useful relationship between ANC of
the Gamow radial wave function and the renormalized (due to the
Coulomb interaction) Coulomb-nuclear partial scattering amplitude.
We use an analytical approximation in the form of a series for the
nonresonant part of the phase shift which can be analytically
continued to the point of an isolated resonance pole in the
complex plane of the momentum. Earlier, this method which we call
the $S$-matrix pole method was used by us to find the resonance
pole energy. We find the corresponding fitting parameters for the
$^5\rm{He},\,^5\rm{Li}$, and $^{16}\rm{O}$ concrete resonance
states.  Additionally, based on the theory of the effective range,
we calculate the parameters of the $p_{3/2}$ and $p_{1/2}$
resonance states of the nuclei $^5\rm{He}$ and $^5\rm{Li}$ and
compare them with the results obtained by the $S$-matrix pole
method. ANC values are found which can be used to calculate the
reaction rate through the $^{16}\rm{O}$ resonances which lie
slightly above the threshold for the $\alpha^{12}\rm{C}$ channel.
\end{abstract}
 \maketitle

\section{Introduction}\label{intr}
It is known that many reactions important for nuclear astrophysics
proceed through subthreshold bound states and lower-lying
resonance states above the threshold and the single-channel
approach can be applied to describe these states. To calculate the
rate of such reactions, we need to find the asymptotic
normalization coefficient (ANC) of the radial wave function for
bound and resonance states. The ANC method has been explored as an
indirect experimental method  for the determination of the cross
sections of  peripheral reactions at low energy \cite{akram2001}.
There are several methods to determine the bound states ANC from
experimental data (see \cite{muk07, akram99} and references
therein). Recently the effective-range expansion method has been
developed to find the ANC for bound and resonant states from an
elastic scattering phase shift analysis (see \cite{orlov, spren}
and references therein). We note that a sufficiently precise
measurement of elastic scattering can give crucial information
concerning the ANC. However, finding the ANC for a resonance is
more difficult than for a bound state. It was shown earlier that
for narrow resonances the ANC is proportional to the square root
of the width $\Gamma$ of the resonance considered \cite{akram77}.
It is known that the normalization procedure for the Gamow wave
function of a resonance, particularly in the case of a broad
resonance when one can not apply the Zel'dovich's formula
\cite{zeld}, is difficult because the outgoing wave increases
exponentially due to the complex momentum.

However, having the ANC, we know the asymptotic part of the wave
function which allows us to normalize it correctly if we choose a
nuclear potential of the interaction between the two nuclei
considered, thus describing the resonant state.

The problem of the exponential increase of the Gamow resonance
wave function in the asymptotic region can be solved by using a
complex scaling method based on the so-called ABC-theorem
\cite{ABC}. This method can be applied to  charged particles as
well (see, for example, \cite{guarmati}) because the Coulomb
potential satisfies the scaling condition of the ABC-theorem. The
complex scaling method using  the Zel'dovich's formula appears
quite widely in the literature (see \cite{akram09} and references
therein). However, the application of this method to a numerical
normalization of the Gamow  wave function is rather difficult. In
\cite{akram09} the problem of calculating the resonance pole was
solved using  similar $S$-matrix pole approach but  for a
potential model, unlike in our present work.

Usually $R$-matrix theory is applied to define the parameters of
low-lying resonances and to describe nuclear resonance reactions.
One of the shortcomings of this theory is the need to fix a value
of the channel radius, which is impossible to measure
experimentally. Therefore, it is important to develop a theory
based on the general properties of the scattering or reaction
amplitudes, which can be used for an analytical continuation to
the nonphysical Riemann energy surface.

We would like to point out that knowing the parameters of
low-lying isolated resonances (in particular the ANC values)
allows us to predict accurately the crucial reaction rates for
nuclear astrophysics.

\section{The ANC from the elastic scattering amplitude based on the analytic properties of the $S$-matrix}\label{S-matrix}
As we mentioned above, the application of the analytic properties
of the $S$-matrix makes it easy to link the ANC to the width
$\Gamma$ of an isolated narrow resonance \cite{akram77}. However,
this relationship is not valid for a wide resonance. In this
section, we show how to obtain this relationship for a resonance
with a broad width as well.

The partial amplitude of the nuclear scattering modified by the
Coulomb interaction is \footnote{Here and below we use the unit
system $\hbar=c=1$.}
\begin{equation}\label{amp}
f_l(k)=\frac{e^{i2\sigma_l}\Bigl(e^{i2\delta_l}-1\Bigr)}{2ik},
\end{equation}
where $k$ is the relative momentum of the colliding nuclei; and
$\delta_l$ is the nuclear scattering phase shift for the orbital
momentum $l$ modified by the Coulomb interaction. (This also
depends on the total angular momentum $J$, which we omit because
Coulomb effects do not depend on the spin.) The $\sigma_l$ is the
pure Coulomb scattering phase shift
\begin{equation}\label{coul-ph}
\sigma_l=\arg\Gamma({l+1+i\eta}),
\end{equation}
or
\begin{equation}\label{gamma}
e^{i2\sigma_l}=\frac{\Gamma(l+1+i\eta)}{\Gamma(l+1-i\eta)},
\end{equation}
where $\Gamma(x)$ is the gamma-function, $\eta=z_1z_2\mu \alpha/k$
is the Sommerfeld parameter, $\alpha$ is the fine-structure
constant, and $\mu$ is the reduced mass of the colliding nuclei
with the charge numbers $z_1$ and $z_2$.

In the single-channel elastic scattering case the partial $S$-matrix
element, without the pure Coulomb part, is
\begin{equation}\label{S-l}
S_l(k)=e^{i2\delta_l}.
\end{equation}
Near an isolated  resonance it can be represented as \cite{migdal}
\begin{equation}\label{S-mat}
S_l(k)=e^{2i\nu_l(k)}\frac{(k+k_r)(k-k_r^\star)}{(k-k_r)(k+k_r^\star)},
\end{equation}
where $k_r=k_0-ik_i$ is the complex wave number of a resonance
 [$k_0>k_i>0$, and the symbol (*) means the
complex conjugate operation]. Energy $E_r$ of this resonance and
its width $\Gamma$ are
\begin{equation}\label{energy res}
E_r=\frac{k_0^2-k_i^2}{2\mu},\qquad \Gamma=\frac{2k_0k_i}{\mu}.
\end{equation}
The partial scattering nonresonant phase shift $\nu_l(k)$ is a
smooth function near the pole of the $S$-matrix element,
corresponding to the resonance. The $S$-matrix element defined by
Eq. (\ref{S-mat}) fulfills the conditions of analyticity,
unitarity, and symmetry. Using Eq. (\ref{S-mat}), one can rewrite
Eq. (\ref{S-l}) in the form
\begin{equation}\label{S-phas}
S_l(k)=e^{2i(\nu_l+\delta_r+\delta_a)},
\end{equation}
where $$\delta_r=-\arctan{\frac{k_i}{k-k_0}}$$ represents the
resonance phase shift, while
$$\delta_a=-\arctan{\frac{k_i}{k+k_0}}$$ is the additional phase
shift which  contributes to the whole scattering phase shift. Thus
the total phase shift is
\begin{equation}\label{phase}
\delta_l=\nu_l+\delta_r+\delta_a.
\end{equation}
The amplitude (\ref{amp}) has a complicated analytical property in
the complex momentum plane due to the Coulomb factor. According to
Refs. \cite{akam84,hamilton,orlov}, we renormalize  the partial
amplitude of the elastic scattering multiplying it by the function
\begin{equation} \label{yost}
h_l(k)=\frac{(l!)^2e^{\pi\eta}}{(\Gamma(l+1+i\eta)^2}
\end{equation}
Applying Eq. (\ref{gamma}),  we can write the renormalized
amplitude as
\begin{equation}\label{tilde-f}
\tilde{f}_l(k)=\frac{(e^{i2\delta_l}-1)}{2ik}\frac{\Gamma(l+1+i\eta)}{\Gamma(l+1-i\eta)}\times
\frac{(l!)^2e^{\pi\eta}}{(\Gamma(l+1+i\eta)^2}.
\end{equation}
After simplification and replacing $e^{i2\delta_l}$ by $S_l(k)$ we
get
\begin{equation}\label{fl}
\tilde{f}_l(k)=\frac{S_l(k)-1}{2ik\rho_l(k)},
\end{equation}
where $\rho_l$ is equal to
\begin{equation}\label{rho}
\rho_l(k)=\frac{2\pi\eta}{e^{2\pi\eta}-1}\prod_{n=1}^l\Bigl(1+\frac{\eta^2}{n^2}\Bigr).
\end{equation}
This renormalized amplitude $\tilde{f}_l(k)$ can be analytically
continued like the partial scattering amplitude, corresponding to
the short-range interaction, and has its pole at  point $k_r$
according to Eq. (\ref{S-mat}). But we should note that the
Coulomb interaction leads to an essential singularity at zero
energy and also (see \cite{IzvRAN2009}) to an infinite number of
poles of  $\tilde{f}_l(k)$ in addition to the poles of a purely
nuclear nature.

In the vicinity of the pole $k_r$, the partial scattering amplitude
(\ref{fl}) can be represented as
\begin{equation}\label{pole}
\tilde{f}_l(k)=\frac{W}{k-k_r}+\tilde{f}_{nonres}(k),
\end{equation}
where the function $\tilde{f}_{nonres}(k)$ is regular at the point
$k_r$.

The simple derivation of the residue $W$ leads to the expression
\begin{equation}\label{residue}
W=\text{res}\tilde{f}_l=\lim_{k\to k_r}\Bigl[(k-k_r)
\tilde{f}_l(k)\Bigr]= -\frac{k_ie^{i2\nu_l(k_r)}}{k_0\rho_l(k_r)}.
\end{equation}
According to the definition of the nuclear vertex constant
$\tilde{G}_l$ (NVC), \cite{blok77}  the relationship between NVC
and the residue $W$ can be written as
\begin{equation}\label{NVC}
W=-\frac{\mu^2}{2\pi k_r}\,\tilde G_l^2.
\end{equation}
So we get
\begin{eqnarray}\label{NVC-2}
\tilde G_l^2&=&\frac{2\pi
}{\mu^2}\frac{k_rk_ie^{i2\nu_l(k_r)}}{k_0\rho_l(k_r)}\nonumber\\
&=&\frac{\pi\Gamma}{\mu
k_0}\frac{(1-ik_i/k_0)e^{i2\nu_l(k_r)}}{\rho_l(k_r)}.
\end{eqnarray}
Using the relationship between NVC $\tilde{G}_l$ and ANC $C_l$
\cite{blok77}, we obtain
\begin{eqnarray}\label{ANC}
C_l&=&\frac{i^{-l}\mu}{\sqrt{\pi}}\frac{\Gamma(l+1+i\eta_r)}{l!}e^{-\frac{\pi\eta_r}{2}}\tilde G_l\qquad\nonumber\\
&=&i^{-l}\sqrt{\frac{\mu\Gamma}{k_0}}e^{-\frac{\pi\eta_r}{2}}\frac{\Gamma(l+1+i\eta_r)}{l!}\nonumber\\
&\times&e^{i\nu_l(k_r)}\sqrt{(1-ik_i/k_0)/\rho_l(k_r)}.\qquad
\end{eqnarray}
The derived equations are valid for both narrow and  broad
resonances. For narrow resonances, when $\Gamma\ll E_r$ ($k_i\ll
k_0$), one can simplify  Eq. (\ref{ANC}) for the ANC replacing
$k_r$ by $k_0$ and using the equality
\begin{equation}\label{simp}
e^{-\frac{\pi\eta}{2}}\frac{\Gamma(l+1+i\eta)}{l!\sqrt{\rho_l(k_0)}}=e^{i\sigma_l}.
\end{equation}
to obtain
\begin{equation}\label{narrow}
C^a_l=\sqrt{\frac{\mu\Gamma}{k_0}}e^{i(\nu_l(k_0)+\sigma_l(k_0)-\pi
l/2)},
\end{equation}
which coincides with the result obtained in Ref. \cite{akram77}.

The nonresonant phase shift $\nu_l(k)$ is the analytical function
excluding the origin. In Ref. \cite{mur83}, the authors presented
the behavior of $\nu_l(k)$ near origin as
\begin{equation}\label{nu-l}
\nu_l(k)=-\frac{2\pi}{(l!)^2}k^{2l+1}\eta^{2l+1}a_le^{-2\pi\eta},
\end{equation}
where $a_l$ is the scattering length for colliding nuclei. We see
that $k=0$ is  an essential singularity of the scattering phase
shift. However, as a function of the momentum $k$, it has normal
analytical properties near the point corresponding to the
resonance. Therefore we can expand $\nu_l(k)$ to a series
\begin{equation}\label{series}
\nu_l(k)=\sum\limits_{n=0}^{\infty}c_n(k-k_s)^n
\end{equation}
in the vicinity of the pole corresponding to the resonance. The
point $k_s$ denotes a centered point, and the radius of
convergence should be shorter than the distance from the centered
point to the closest singular point. The last can be due to an
exchange Feynman diagram for the elastic  scattering, leading to
the logarithmic singularity which is absent in our model.

If we wish to determine the value of the phase shift $\nu_l(k)$ by
applying Eq. (\ref{series}) at a point on the complex plane close
to the centered point $k_s$, then  only the first few items of the
convergent series for calculating $\nu_l(k)$ can be taken into
account with certain precision.

The expansion coefficients $c_n$ of Eq. (\ref{series}) as well as
$k_0$ and $k_i$ are determined by  fitting the experimental values
of the elastic scattering phase shifts $\delta_l$ given by Eq.
(\ref{phase}).

\section{The effective-range method}\label{eff-range}
The effective-range theory is also based on the analytical
property of the elastic scattering amplitude when an ingoing
particle collides with another nuclei at low energy. This is a
a very good method to find the NVC and ANC of the bound states from
phase shift analyses (see Refs. \cite{orlov}, \cite{spren} and
references therein).

Substituting the expression Eq. (\ref{S-l}) of the partial
$S$-matrix into  Eq. (\ref{fl}) we easily obtain the renormalized
amplitude in the following form:
\begin{equation}\label{fl2}
\tilde{f}_l(k)=\frac{1}{k(\cot\delta_l-i)\rho_l(k)},
\end{equation}
where the function $\rho(k)$ is defined by Eq. (\ref{rho}) and
$\delta_l$ is the nuclear phase shift modified by the Coulomb
interaction. From Eq. (\ref{fl2}) it follows that the position of
the pole corresponds to the condition
\begin{equation}\label{zero}
\cot\delta_l-i=0.
\end{equation}
Exactly the same condition (\ref{zero}) is fulfilled for the pole
of the elastic scattering amplitude of the uncharged particles.
Following Ref. \cite{haer} we write the effective-range function,
which is an analytical function, except for possible poles (zeros
of the scattering amplitude), and relates to the phase shift
$\delta_l$ as
\begin{equation} \label{CoulombKl}
K_l(k^2) = k^{2l+1} D_l(\eta)\left[C_0^2(\eta)(\cot\delta_l - i)+
2 \eta h(\eta )\right],
\end{equation}
where
\begin{equation}
C_0^2(\eta)=\frac{2\pi \eta} {\exp(2\pi\eta)-1},
\end{equation}
\begin{equation} \label{H-eta-sigma}
h(\eta)= \psi(i\eta) + (2i\eta )^{- 1} - \ln(i\eta),
\end{equation}
\begin{equation} \label{Dl-eta}
D_l(\eta)=\prod_{n=1}^l(1+\eta^2/n^2),\quad D_0(\eta)=1,
\end{equation}
and $\psi(x)$ is the digamma function. We note that the
effective-range function  $K_l(k^2)$ is real in the positive
energy region.

If the interaction of colliding particles is purely nuclear, i.e.
without the Coulomb tail, the effective-range function
(\ref{CoulombKl}) is simplified and expressed through the partial
scattering phase shift by the well known equation
\begin{equation}\label{K-nuclear}
K_l(k^2)=k^{2l+1}\cot\delta_l.
\end{equation}
Since the effective-range function is an analytic function(except
for possible poles), it can be expanded in a power series over
$k^2$ in the low energy region, where only the elastic scattering
channel is open.
 Typically, the following expansion is used
\begin{equation}\label{ef-expand}
K_l(k^2)=-\frac{1}{a_l}+\frac{1}{2}r_l^2k^2-P_lr_l^3k^4+\cdots,
\end{equation}
where $a_l$, $r_l$ and $P_l$ are real and called the scattering
length, effective range and shape parameter, respectively. An
alternative form to (\ref{ef-expand}) is the Pad\'e-approximation
used in Ref. \cite{Pade}.

The expansion coefficients of Eq. (\ref{ef-expand}) are defined by
fitting the effective-range function expressed through
experimental phase shifts for the positive energy  in the form of
Eq. (\ref{CoulombKl}) or Eq. (\ref{K-nuclear}), depending on
whether a charged or uncharged particle is scattered by the target
nucleus. The effective-range function Eq. (\ref{ef-expand}) with
the fitted parameters is used to find the pole of the elastic
scattering amplitude, corresponding to the condition of
(\ref{zero}) which leads to the equation
\begin{equation}\label{cond-zero}
K_l(k^2)-2\eta k^{2l+1}D_l(\eta)h(\eta)=0.
\end{equation}

Actually, Eq. (\ref{cond-zero}) can be taken as a condition for
parameter fitting when a resonance pole energy and a width are
included as an input like the phase shift data.

For the pole of the elastic scattering amplitude in the case of an
uncharged particle, the pole condition is simplified to
\begin{equation}\label{uncharged-pole}
K_l(k^2)-i k^{2l+1}=0.
\end{equation}
Solving Eq. (\ref{cond-zero}) or (\ref{uncharged-pole}), we find
the pole momentum value of the elastic scattering amplitude and
the energy which has complex value for a resonance respectively.
Then we calculate the residue $W$ of the renormalized scattering
amplitude of a charged particle (\ref{fl2}) at this pole point.
The equation for $W$ is
\begin{equation} \label{res1}
W =\frac{k^{2l}}{\frac{d}{dk}\left[K_l(k^2)-2 \eta
k^{2l+1}D_l(\eta)h(\eta)\right]}\mid_{k=k_r},
\end{equation}
for a charged particle and
\begin{equation} \label{res2}
W
=\frac{k^{2l}}{\frac{d}{dk}\left[K_l(k^2)-ik^{2l+1}\right]}\mid_{k=k_r},
\end{equation}
in the case of an uncharged particle scattering.

The expressions for the NVC and ANC are defined through the
residue $W$ by Eqs. (\ref{NVC}) and (\ref{ANC}), which are given
in the previous section.

\section{Results for the $^5\rm{He}$ and $^5\rm{Li}$ ground and first exited
states}\label{N-alpha}

The $^5\rm{Li}$ and $^5\rm{He}$ nuclei are interesting in that the
ground and first excited states are resonance states which can be
treated as single-channel systems.  The phase shift of the elastic
$N\alpha$ scattering with total angular momentum and parity equal
to $J^\pi=3/2^{-}$ passes rapidly through $\pi/2$ and therefore
leads to a narrow resonance. However, the phase shift of the
elastic $N\alpha$ scattering with $J^\pi=1/2^{-}$ does not pass
through $\pi/2$ and therefore the corresponding resonance  is wide
enough. This fact leads to certain difficulties, not only in
determining the position of the resonance and its width, but also
in finding such characteristics as the NVC and ANC.

The coefficient values of the effective-range expansion obtained
from a phase shift analysis of the elastic scattering data in the
region up to 3 MeV for neutron and 5 MeV for proton were found by
the authors of Ref. \cite{ardnt}. Using these parameters, the
authors of Ref. \cite{ahmed} determined the values of the energy
and width of the resonances. The article \cite{ahmed}  was cited
in Ref. \cite{lechman} where a separable potential fits the
resonance parameters for the $n\alpha$ scattering in the $p_{1/2}$
and $p_{3/2}$ states. Agreement of the phase shifts calculated in
\cite{lechman} with the experimental ones is good for the narrow
$p_{3/2}$ resonance but is poor for the broad $p_{1/2}$ resonance.
The N/D method was applied in \cite{akam84,safron} for calculating
the values of the parameters of these resonances. Additionally,
the residues $W$ of the renormalized scattering amplitude were
calculated at the resonance poles in the complex $k$ plane using
the effective-range method in \cite{orlov}.

\begin{figure*}[thb]
\begin{center}
\parbox{8.0cm}{\includegraphics[width=8.0cm]{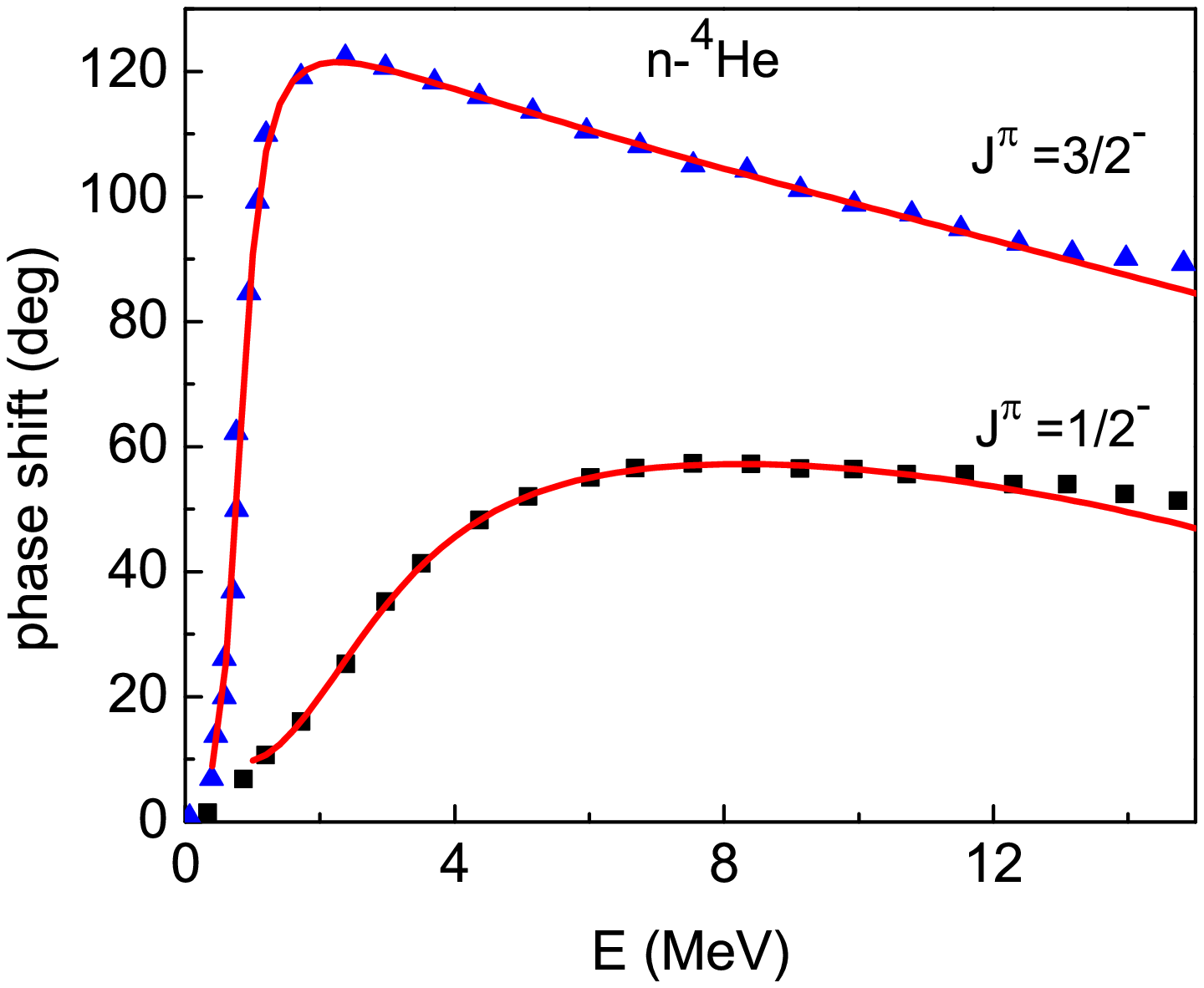}}
\hspace*{1cm}
\parbox{8.0cm}{\includegraphics[width=8.0cm]{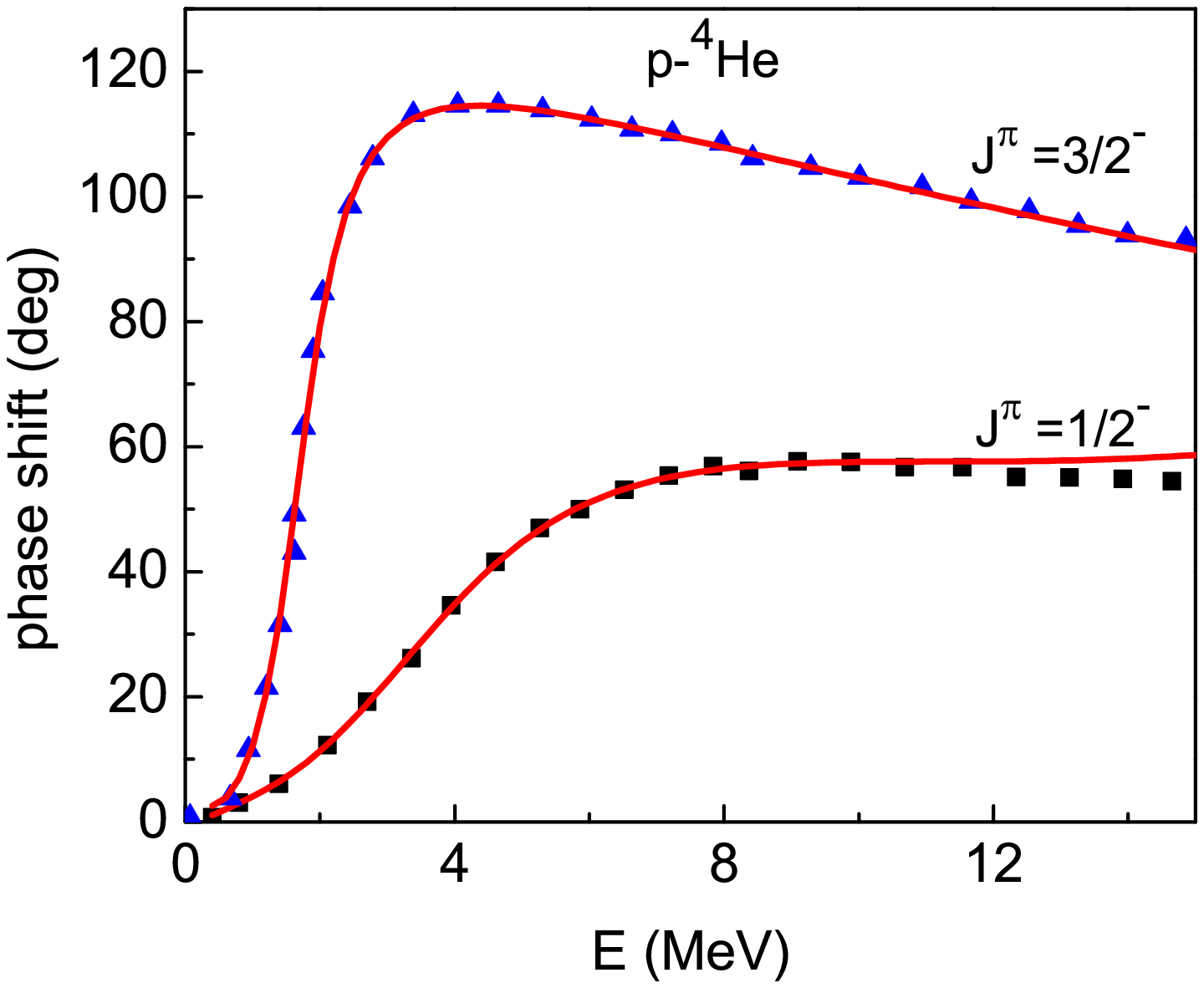}}
\end{center}
\caption{(Color online)  Comparison of the fitted phase shifts for
the $n(p)$ -$^4\rm{He}$ elastic scattering obtained by the
$S$-matrix pole method  with the experimental values. The
experimental data are taken from Ref. \cite{forssen}. The energy
is given in the laboratory frame.  \label{fig1}}
\end{figure*}

We applied the $N\alpha$ phase shifts data presented in Ref.
\cite{forssen} to calculate $W$, NVC and ANC. According to the
authors of Ref. \cite{forssen} the $N\alpha$ phase shifts are
obtained by an accurate $R$-matrix analysis of the elastic
scattering data. In Fig. \ref{fig1} we show the results of fitting
the phase shifts  for the $n$ -$^4\rm{He}$ and $p$ -$^4\rm{He}$
elastic scattering, using the $S$-matrix pole method. A good
agreement  is achieved in the wide energy region, including the
resonances considered.

In Table \ref{tab1} we present the parameter values related to the
$^5\rm{He}$ and $^5\rm{Li}$ nuclei, which are calculated using the
analytic properties of the $S$-matrix outlined in Sec.
 \ref{S-matrix}.
\begin{table*}[!ht]
\caption{Nucleus, channel, state, energy and width, corresponding
values of the residue ($\mid W\mid$),  NVC ($\tilde{G}^2_l$) and
ANC ($C_l$) obtained by fitting the elastic $N\alpha$ scattering
phase shifts presented in Ref. \cite{forssen}. Results are found
using the analytical properties of the $S$-matrix outlined in Sec.
\ref{S-matrix}. Four terms of Eq. (\ref{series}) are used for
fitting. The last column shows the ANC ($C_l^a$) calculated by Eq.
(\ref{narrow}). The energy of the resonance is given in the
center-of-mass system of $N\alpha$.}
\begin{ruledtabular}
\begin{tabular}{lccccrrr}
&&&&&&&\\
Nucleus&$J^\pi$&$E_r$\,(MeV)&$\Gamma$\,(MeV)&$\mid W\mid$
&$\tilde{G}^2_l$\,(fm)\quad\,\,&$C_l\,(\rm{fm}^{-1/2})$&$C^a_l\,(\rm{fm}^{-1/2}$)\\
&&&&&&&\\
\hline
&&&&&&&\\
$^5\rm{He};\,n\alpha$&$3/2^-$&0.629& 0.448 &0.147&$0.005- i0.009$&$-0.105-i0.190$&$-0.095-i0.214$\\
&$1/2^-$&1.476& 3.520&0.194&$-0.019-i0.016$&$-0.320-i0.116$&$-0.391-i0.314$\\
$^5\rm{Li};\,p\alpha$&$3/2^-$&1.328& 0.994&0.320&$0.018-i0.027$&$-0.115-i0.231$&$-0.103-i0.269$\\
&$1/2^-$&2.504& 4.667&0.261&$-0.011-i0.040$&$-0.276-i0.196$&$-0.355-i0.374$\\
 \end{tabular}
 \end{ruledtabular}
\label{tab1}
\end{table*}

\begin{figure*}[thb]
\begin{center}
\parbox{8.0cm}{\includegraphics[width=8.0cm]{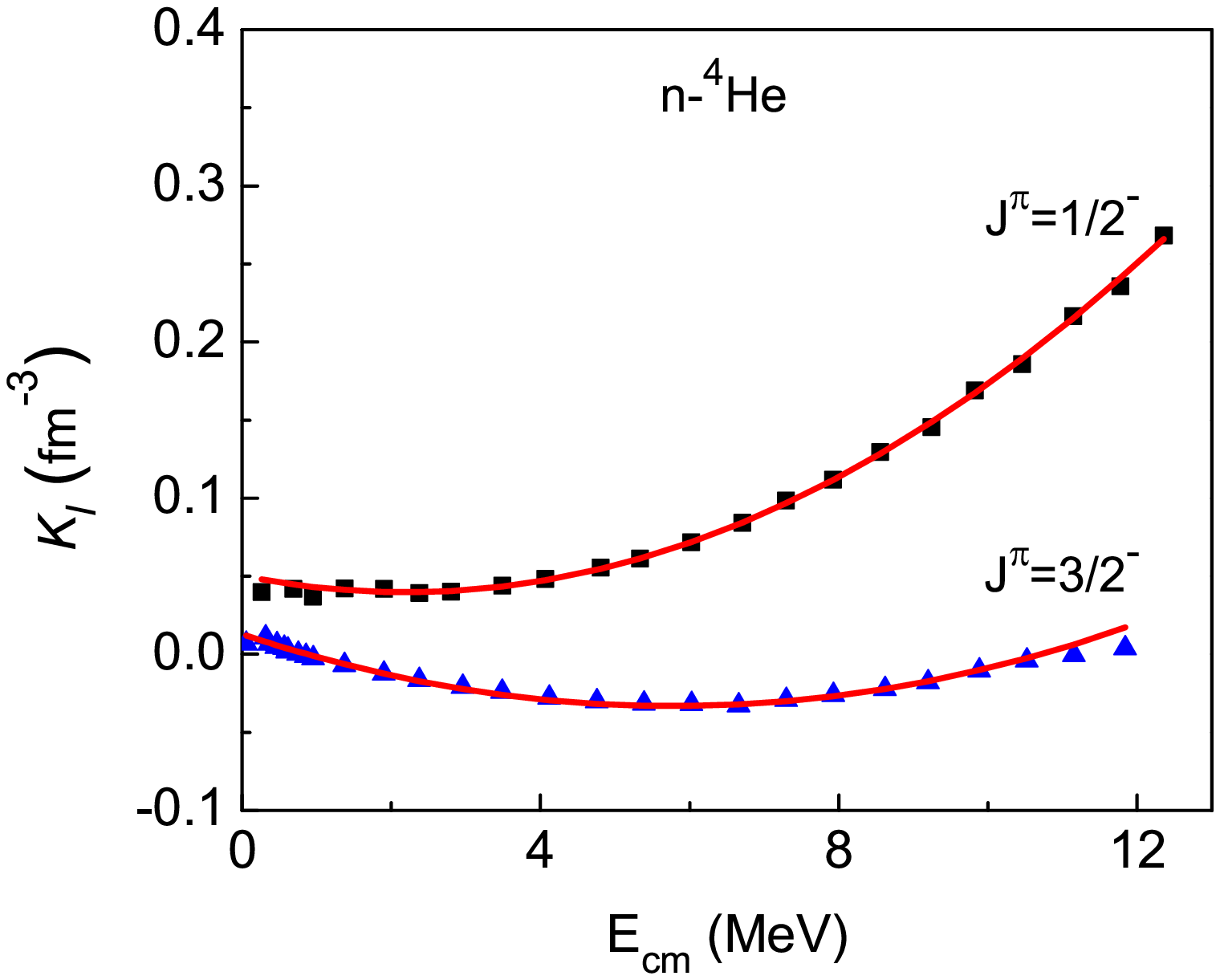}}
\hspace*{1cm}
\parbox{8.0cm}{\includegraphics[width=8.0cm]{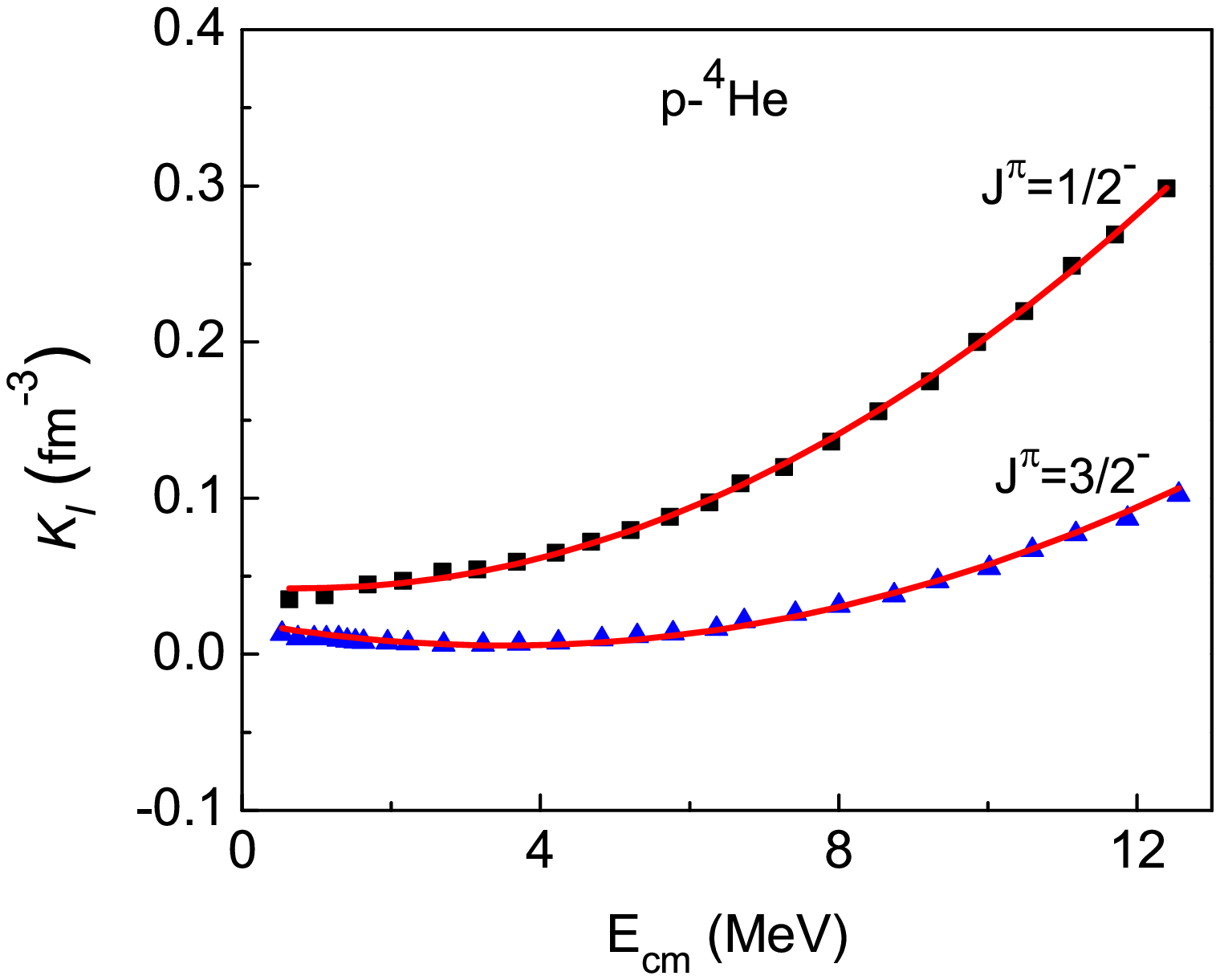}}
\end{center}
\caption{(Color online)  Comparison of the fitted effective-range
functions for the $n(p)$-$^4\rm{He}$ elastic scattering with  the
experimental values calculated by using
 the experimental data  taken from Ref. \cite{forssen}.  The energy is given in the c.m. frame.  \label{fig2}}
\end{figure*}

In Fig. \ref{fig2} we  compare the fitted effective-range function
with the corresponding values calculated by the effective-range
method, using the experimental phase shift data taken from
\cite{forssen}. The obtained agreement is quite good.

 Table \ref{tab2} shows the  calculation results of the same
parameters for the same nuclei and states, but found using the
effective-range method described in Sec. \ref{eff-range}.
\begin{table*}[!ht]
\caption{Same as in Table \ref{tab1}, but in the frame of the
effective-range method outlined in Sec. \ref{eff-range}.}
\begin{ruledtabular}
\begin{tabular}{lccccrrr}
&&&&&&&\\
Nucleus&$J^\pi$&$E_r$\,(MeV)&$\Gamma$\,(MeV)&$\mid W\mid$
&$\tilde{G}^2_l$\,(fm)\quad\,\,&$C_l\,(\rm{fm}^{-1/2})$&$C^a_l\,(\rm{fm}^{-1/2}$)\\
&&&&&&&\\
\hline
&&&&&&&\\
$^5\rm{He};\,n\alpha$&$3/2^-$&0.675& 0.560 &0.171&$0.007- i0.010$&$-0.111-i0.212$&$-0.076-i0.245$\\
&$1/2^-$&1.563& 4.155&0.220&$-0.015-i0.026$&$-0.323-i0.187$&$-0.384-i0.367$\\
$^5\rm{Li};\,p\alpha$&$3/2^-$&1.481& 1.041&0.295&$0.019-i0.025$&$-0.109-i0.236$&$-0.062-i0.281$\\
&$1/2^-$&2.213& 4.640&0.305&$-0.016-i0.043$&$-0.300-i0.193$&$-0.375-i0.369$\\
 \end{tabular}
 \end{ruledtabular}
\label{tab2}
\end{table*}

A comparison of the results presented in Tables \ref{tab1} and
\ref{tab2} shows that both methods lead to quite consistent
results. The essential difference between some of the results for
the two methods considered may be explained by the fact that these
results are more sensitive to the applied approach in the case of
broad resonances. The same conclusion was noted in Ref.
\cite{csoto}, where the authors also analyzed the parameters of
the $N\alpha$ states given in Refs. \cite{bond,schw}. We would
like to point out that the difference between the energies of
states $1/2^-$ and $3/2^-$ for $^5\rm{He}$ received by both
methods applied is  $\sim 0.9$ MeV, which is comparable to the
difference $\sim 1.1-1.3$ MeV between results given by other
authors. (See the tables in \cite{orlov, csoto}). The same
differences for the states $^5\rm{Li}$ are 1.17 and 0.71 MeV,
which are obtained using the presentation of the $S$-matrix [Eq.
(\ref{S-mat})] and the effective-range method, respectively. The
results found by the other authors lead to values where the limits
are relatively wide. As to the widths of the corresponding levels,
the range of differences of the values obtained by the different
authors is similar to the range of differences for the real parts
of the resonance energies \cite{orlov, csoto}.

According to our results, the difference in the level energies
 calculated by the two methods described above, are 6-7\% for
$^5\rm{He}$ and 11\% for $^5\rm{Li}$, while the width differences
are 7\% at state $J^\pi=3/2$ and 18\% for $J^\pi=1/2$ of
$^5\rm{He}$. For the levels of $^5\rm{Li}$, differences in the
widths calculated by the two methods are very small. Comparing the
results of Tables \ref{tab1} and \ref{tab2} we can see that most
of the calculated data have similar values with a maximum
difference of $\sim 20\%$. From this comparison it can be
concluded that it is difficult to decide which method of
calculation is preferable. Comparing the values of the ANC of the
penultimate and last columns, we see a difference in $\sim 60\%$,
which gives us grounds to say that the asymptotic formula defined
by Eq. (\ref{narrow}) leads to incorrect values of the ANC for
broad resonances.  The values of the residue [Eqs.(\ref{residue})
and (\ref{res1})] calculated by both methods are similar in
absolute values to the corresponding values presented in Refs
\cite{safron,orlov}.

\section{Results for the $^{16}\rm{O}$  low-lying resonances situated
above the $\alpha^{12}\rm{C}$ threshold} \label{alpha-12C}

\begin{figure}[thb]
\begin{center}
\parbox{8.0cm}{\includegraphics[width=8.0cm]{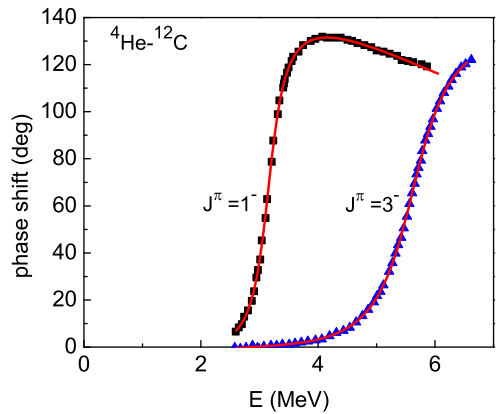}}
\end{center}
\caption{(Color online)  Comparison of the fitted phase shifts for
the  $\alpha$ -$^{12}\rm{C}$ elastic scattering obtained by the
$S$-matrix pole method with the experimental values. The
experimental data are taken from Ref. \cite{tisch}. The energy is
given in the laboratory frame.  \label{fig3}}
\end{figure}

 In our previous work \cite{blok11}, we determined the position and the
width of the resonance in $^{16}\rm{O}$, using Eqs. (\ref{phase})
and (\ref{series}) by fitting the phase shift for the elastic
scattering of the $\alpha$ particles on the nucleus $^{12}\rm{C}$
given in Ref. \cite{tisch}. It was found that the dependence of
the results on the location of $k_s$ is insignificant if it is
within the area of the maximum increase of the full scattering
phase shift. To verify the almost linear behavior of the phase
shift $\nu_l(k)$, we checked its dependence on the momentum $k$ by
subtracting the sum of the phase shifts $\delta_r(k)$ and
$\delta_a(k)$ from the experimental phase shift within the
resonance region. In Fig. \ref{fig3} we demonstrate a good
description of the energy dependence of the experimental
$\alpha^{12}\rm{C}$ elastic scattering phase shifts which is
obtained using the $S$-matrix pole method.  As examples, we take
the
 $J^\pi=1^-$ and $J^\pi=3^-$ states,when the resonances are broad enough.

We note that all known methods of fitting the elastic scattering
phase shift lead to the same values of the energy and width for
narrow resonances. However, the results diverge for broad
resonances. Therefore, we can expect a difference in the results
of the ANC evaluations for broad resonances compared with
calculations by Eq. (\ref{narrow}).  Table \ref{tab3} shows our
calculation results for the energy and width of the resonances for
the nucleus $^{16}\rm{O}$, and the corresponding NVC and ANC
values.
%\begin{widetext}
\begin{table*}[!ht]
\caption{States, energies and widths of $^{16}\rm{O}$ nucleus
levels above the $\alpha^{12}\rm{C}$ threshold from our fit, as
well as the corresponding values of the calculated NVC and ANC
from the elastic $\alpha^{12}\rm{C}$ scattering phase shifts
\cite{tisch}. Four terms of Eq. (\ref{series}) are used for
fitting. The energies of the resonances are given in the
center-of-mass system of $\alpha^{12}\rm{C}$}.
\begin{ruledtabular}
\begin{tabular}{lccccrrr}
&&&&&&&\\
$J^\pi$&$E_r$\,(MeV)\cite{tisch}&$\Gamma$\,(keV)\cite{tisch}&$E_r$\,(MeV)&$\Gamma$\,(keV)
&$\tilde{G}^2_l$\,(fm)\,\,\,\,\,\,\,\,&$C_l\,(\rm{fm}^{-1/2})$\,\,\,&$C^a_l\,(\rm{fm}^{-1/2}$)\,\,\,\,\\
&&&&&&&\\
\hline
&&&&&&&\\
$0^+$&4.887& 3.0 &4.887& 3.0&$0.0023-i 0.0042$&$0.0122- i 0.0104$&$0.0122- i0.0104$\\
$1^-$&2.416& 388.0&2.364& 356.2&$4.9703-i 1.7969$&$0.1530-i 0.1032$&$0.1759- i 0.1135$\\
$2^+$&2.683& 0.76&2.683&0.76&$0.0031-i 0.0002$&$0.0038- i0.0086$&$0.0038- i 0.0086$\\
$2^+$&4.339& 83.0&4.350&79.1&$0.0383-i 0.0079$&$-0.0125-i0.0831$&$-0.0124-i 0.0838$\\
$3^-$&4.320&864.0&4.214&811.7&$0.2762-i 0.1420$&$-0.2332-i 0.0201$&$-0.2718- i 0.0311$\\
$4^+$&3.196&25.6&3.199&26.5&$0.0284-i 0.0014$&$-0.0491+ i 0.0190$&$-0.0494+ i 0.0190$\\
 \end{tabular}
 \end{ruledtabular}
\label{tab3}
\end{table*}
%\end{widetext}
In the second and third columns of Table \ref{tab3} we show the
results obtained by a $R$-matrix analysis \cite{tisch} while our
results received by a $S$-matrix analysis are displayed in the
fourth and fifth columns. Readers can see that these results for
the energy and width coincide when the resonance is narrow, but
there are essential differences for broad resonances (in
particular for states $1^-$ and $3^-$). The values of the
renormalized NVCs ($\tilde{G}^2_l$) and ANCs ($C_l$), which were
found by using our calculated values of the energies, widths and
the nonresonant phase shifts are shown in the next two columns. In
the last column the values of  ANCs ($C^a_l$) which were
calculated by using Eq. (\ref{narrow}), are presented. We note
that these values are found at real momentum values.

As the experimental phase shifts are determined with some
uncertainties, it is reasonable to assess the change of NVC and
ANC as functions of the resonance energy and width. Therefore, we
calculated the value of the nonresonant phase shift $\nu_l$ and
found the values of NVC and ANC at the resonance point for the
state $J^\pi=3^-$, fixing the resonance energy and width fitted by
$R$-matrix method \cite{tisch}. It was found that the differences
in energy and resonance were  2.5\% and 6.4\% respectively, while
the renormalized NVC and ANC differ by 2.9\% and 5.3\%
respectively. It should be noted that the percentage difference of
the NVC and ANC values is a consequence of the calculation of ANC
through the value NVC, because it is multiplied by the $\Gamma(x)$
function at the different values of the Coulomb factor. One can
see that the uncertainties of NVC and ANC are roughly the same as
those of the resonance energy value. For narrow resonances, it is
quite reasonable to evaluate ANC using Eq. (\ref{narrow}), taking
the value of the nonresonant phase shift for the real values of
energy or momentum from the experimental data. It is obvious that
for broad resonances the width of which is greater than their
energy, the uncertainty of the ANC value should be related to  the
uncertainty of the width which is determined by fitting the
experimental scattering phase shifts.

The effective-range method is not able to reproduce the widths of
the $^{16}\rm{O}$ resonances. This may be due to the
single-channel approximation which we use in this work.
\section{Conclusion}
The $S$-matrix pole prescription [ Eq. (\ref{S-mat})] and
expansion of the nonresonant phase to series [Eq. (\ref {series})]
give consistent resonance parameters for the ground and first
excited states of $^5\rm{He}$ and $^5\rm{Li}$  as well as for the
low-lying states of $^{16}\rm{O}$ situated above the
$\alpha^{12}\rm{C}$ threshold in spite of their resonance widths.

The standard expansion of the effective-range function $K_l(k^2)$
to find the NVC $\tilde{G}_l$  and other parameters of the two
first resonance states of $^5\rm{He}$ and $^5\rm{Li}$ are used
successfully. We have found results which are a little different
from those obtained by other methods used. In our opinion, these
differences can be explained by the fact that in the first method,
a centered point of the expansion of the nonresonant phase shift
to a series is the point which is closest to the position of the
resonance, while in the method of the effective range, we use a
centered point of the expansion at zero momentum, which is far
from the resonance pole.

In the case of a bound state, the binding energy can be considered
as an additional parameter unlike in the $S$-matrix method with
the phase shift fitting. Therefore, we expect that the method
using the $S$-matrix pole prescription [Eq. (\ref{S-mat})] can
lead to quite different results, and so we recommend using the
$S$-pole prescription to specify resonance parameters. At the same
time, the effective-range expansion method in the convergence
energy region is applicable in the case of a bound state when the
the $S$-matrix pole prescription does not work.

The results of this paper can be used for solving  nuclear
astrophysical problems and may be applied in the theory of nuclear
reactions using Feynman diagrams to describe the reaction
mechanisms.

This work was supported by the Russian Foundation for Basic
Research (project No. No. 13-02-00399). We are grateful to Ms.
H.\,M.~Jones for editing the English of this manuscript.

\end{document}